**Contributor's note:**

*This is the accepted version of the article*

Link to the final version:

https://onlinelibrary.wiley.com/doi/full/10.1002/pssr.201900381

# Imaging Nanoscale Inhomogeneities and Edge Delamination in As-Grown $MoS_2$ Using Tip-Enhanced Photoluminescence

*Alvaro Rodriguez, Tim Verhagen, Martin Kalbac, Jana Vejpravova, Otakar Frank**

Dr. A. Rodriguez, Dr. M. Kalbac, Dr. O. Frank
J. Heyrovský Institute of Physical Chemistry, Academy of Sciences of the Czech Republic, Dolejškova 2155/3, 182 23 Prague, Czech Republic
E-mail: otakar.frank@jh-inst.cas.cz

Dr. T. Verhagen, Dr. J. Vejpravova
Department of Condensed Matter Physics, Faculty of Mathematics and Physics, Charles University, Ke Karlovu 5, 121 16, Prague 2, Czech Republic

Methods for nanoscale material characterization are in ever-increasing demand, especially those that can provide a broader range of information at once. Near-field techniques based on combinations of scanning probe microscopy (SPM) and Raman or photoluminescence (PL) spectroscopy (tip-enhanced Raman spectroscopy [TERS] and/or tip-enhanced PL [TEPL]) are, thanks to their capabilities and fast development, strong candidates for becoming widespread across the scientific community as SPM and Raman microscopy did only a decade or two ago. The present work describes a gap-less TEPL study performed directly on as-grown $MoS_2$ monolayer samples without any pretreatment or transfer, i.e. without the utilization of plasmonic substrate. Thanks to a mapping resolution as low as a few tens of nanometers, homogeneous layer interiors from defective edge fronts in the grown monolayers could be distinguished. With the aid of additional high-resolution SPM modes, like local surface potential and capacitance measurements, together with nanomechanical mapping, a combination of defects and a lack of substrate doping is suggested as being responsible for the observed PL behavior in the partially delaminated $MoS_2$ layers. In contrast, mechanically exfoliated flakes show topography- and contamination-related heterogeneities in the whole flake area.

Standard optical techniques, such as Raman or photoluminescence (PL) microspectroscopies, give plentiful information about both the structural and optoelectronic properties of layered transition metal dichalcogenides (TMD).[1] However, the spatial resolution of these techniques is diffraction-limited, in which the Abbe diffraction limit of the resolution, $d$, is equal to $d = \lambda/2$ NA, with $\lambda$ the being the wavelength of the irradiation and NA the numerical aperture of the objective. For a typical Raman or PL experiment, the utilization of visible light and an objective with an NA of around 1 result in a diffraction-limited resolution of several hundreds of nanometers, which hinders the understanding of the material's properties at the nanoscale. One possibility for overcoming this issue is the use of the apex of an atomic force microscope (AFM) or scanning tunneling microscope tip (nanoantenna) to improve the intensity of the optical signal in the near-field using the combination of a lightning rod effect and a surface plasmon resonance at the metallized tip.[2] In general, tip-enhanced techniques have greatly improved over the last few years, reaching a resolution of around tens of nanometers, with enhancement factors of $10^7$–$10^{10}$ in the near-field.[3] Even so, most works deal with single spectrum acquisition and only in the past few years have studies concerning tip-enhanced imaging methods been seen, e.g. ref. [4].

TMDs usually exhibit a strong PL associated with an exciton transition. In particular, an $MoS_2$ monolayer presents two direct optical transitions of the A and B excitons at approximately 1.8 eV and 2 eV, respectively. The transition energy can be readily shifted toward higher or lower energies depending on environmental conditions.[5] For instance, the PL of $MoS_2$ monolayers created through chemical vapor deposition (CVD) have been found to shift toward lower energies after transferring to a different substrate with a polymer, due to defects generated during the transfer.[6] Moreover, one can find a great variety of PL energy values reported in the literature due to the diverse synthesis or exfoliation methods used.[7] Nanoscale heterogeneities, such as defects and wrinkles, are most likely responsible for the reported variations, as was shown for graphene previously.[8] In this sense, the, at a minimum, 10- to 20-fold improvement in spatial resolution of tip-enhanced methods can help us to understand how nanoscale heterogeneities may affect the optoelectronics properties of these materials and eventually improve the synthesis methods to obtain optimal CVD flakes for a specific application. Some work on Raman nanoscale characterization of TMD materials has been reported, most obtained in the so-called “gap mode” in which the layers are first transferred to gold substrates to localize and intensify the near-field between the tip and the substrate.[9] However, while both the enhancement and spatial resolution are improved, the transfer inevitably leads to damage in the interrogated material. The obtained information on the quality and properties of the grown layers has thus only a limited value. Moreover, PL is quenched when $MoS_2$ is supported on freshly prepared gold,[10] thereby removing the option of measuring tip-enhanced photoluminescence (TEPL) under this favorable condition.

Here, we report a study of CVD-grown $MoS_2$ flakes using TEPL in the “gap-less” regime, i.e. where the near-field is produced solely by the tip. We measured TEPL of as-grown $MoS_2$ on $SiO_2$/Si substrates with no transfer or other treatment. Different flakes were examined to verify the heterogeneities of the flakes, such as variations of PL at the nanoscale due to defects and the substrate. The results are compared to data obtained on mechanically exfoliated $MoS_2$, with special attention paid to the edges of the flakes.

**Figure 1a** shows a topography image of an $MoS_2$ flake grown by CVD, which reveals significantly taller edges than rather flat interior of the flake. In the height profile plotted in Figure 1b, a thickness of approximately 0.7 nm can be observed, which is consistent with the thickness reported for a CVD-grown monolayer.[11] A remarkable effect can be noticed when analyzing the edges in detail. As observed in Figure 1c, the edges exhibit irregular profiles, with different heights varying in a range of 3 nm. This fluctuation is caused by small particles aligned along the edge of the flake. These nanoparticles are comparable to those observed during the growth of CVD $MoSe_2$, which have been attributed to $MoO_xSe_y$ residues.[12]

Figure 1d and 1e show the Raman and PL spectra measured at the center of the flake. The Raman spectrum features the $MoS_2$ normal modes E' and $A'_1$ separated by less than 21 $cm^{-1}$, which is a characteristic value for CVD-grown monolayer $MoS_2$.[4h] Concerning the PL spectrum, the conventional emissions belonging to the A and B excitons appear at 1.81 eV and 1.96 eV, respectively, confirming the monolayer's thickness. Figure 1f shows the contact potential difference (CPD) map of the flake measured by frequency-modulated Kelvin probe force microscopy (KPFM-FM). In a KPFM-FM measurement, the CPD signal is collected together with the $\partial^2C/\partial z^2$ magnitude, which is directly proportional to the relative change in capacitance and is consequently measured in arbitrary units (Figure 1g).[4g] KPFM measures the CPD between the tip and the sample and provides information about the work function with nanoscale resolution. In $MoS_2$ films, it has been reported that both contaminants and surface defects can modify the CPD and thus modify the work function by up to 150 meV.[13] Both the CPD and $\partial^2C/\partial z^2$ images show narrowly spread values in the interior of the flake, while variations appear only at the edges, correlating well with the topography image (Figure 1a). CPD and $\partial^2C/\partial z^2$ cross-sections across the flake are plotted in **Figure 1h and 1i**, respectively. The CPD inside the flake is approximately 400–500 mV lower than the substrate. The edge shows CPD values that are lower, by an additional 100 mV, than the flake interior (Figure 1h). No noticeable variations in either CPD or $\partial^2C/\partial z^2$ magnitude is detected within the edge region, most likely due to the lower resolution of the KPFM than in topography.[14]

At this point, it was beneficial to gather information about the PL with a similar resolution to the AFM images. As noted above, micro PL resolution is limited to hundreds of nanometers, and one approach to overcoming the diffraction limit is by means of TEPL measurements.

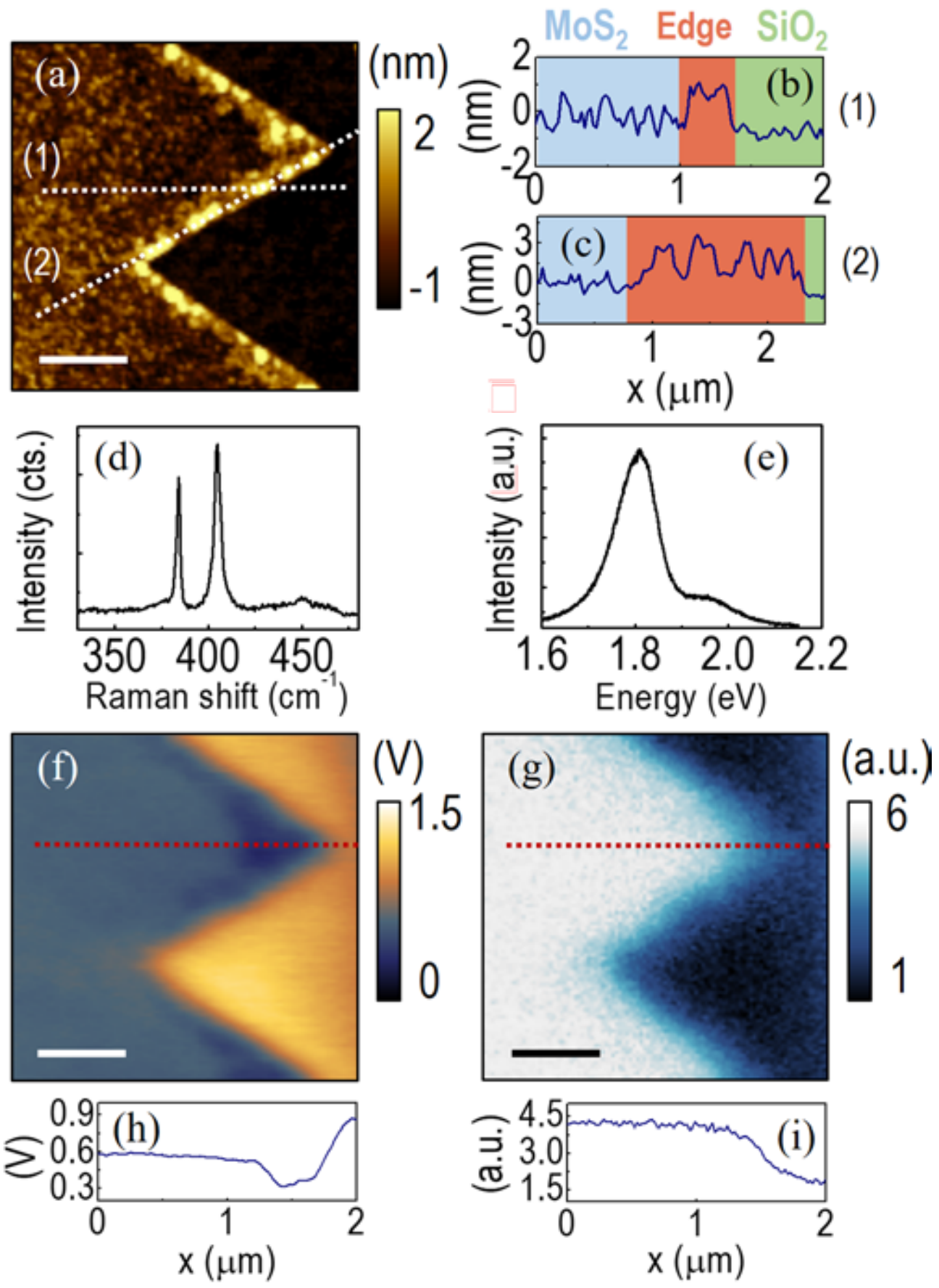

**Figure 1.** CVD-grown $MoS_2$ monolayer: (a) topography image and height profiles (b) across and (c) along the edge. (d) Raman spectrum and (e) photoluminescence spectra acquired in the flake interior. (f) CPD and (g) $\partial^2C/\partial z^2$ maps of the same area with (h) CPD and (i) $\partial^2C/\partial z^2$ profiles across the edge. Scale bar in each map is 500 nm.

We performed TEPL measurements by side illumination of a silver-coated AFM tip with a red laser excitation (633 nm). Far- and near-field images were collected by switching the AFM to tapping and contact mode, respectively, while scanning the sample in tapping mode and measuring the PL spectrum at each point when the tip was close to the sample (near-field spectrum) and far from the sample (far-field spectrum). Side illumination resulted in a rather strong far-field and a non-circular laser spot at the sample.[15] To obtain the sole near-field contribution, subtraction of the far-field needed to be performed. In the following text, we denote the as-measured near-field spectra as "uncorrected near-field," and the corrected near-field data simply as "near-field." In each map point, both far- and near-field PL spectra were fitted to a single asymmetric pseudo-Voight function to account for the superposition of both the neutral ($A^0$) and charged exciton ($A^-$, trion), which are difficult to deconvolute during high-speed TEPL mapping. **Figure 2a** shows the uncorrected near-field (blue line), far-field (black line), and near-field PL spectra (red line) collected in the interior area of the flake. The spectra with the same color coding in Figure 2b correspond to the edges of the flake, where the PL enhancement is higher. Figure 2c and 2d shows maps of the PL intensity and the energy, respectively, of the A exciton in the near-field conditions. PL intensity and energy values did not vary significantly in the interior of the flake. Only the edges, approximately 80 nm wide, showed a prominent change in the PL. The recognition of the features at the edges is possible only thanks to the dramatic improvement of the lateral resolution over the far-field measurements (*cf.* Figure S1, Supporting Information).

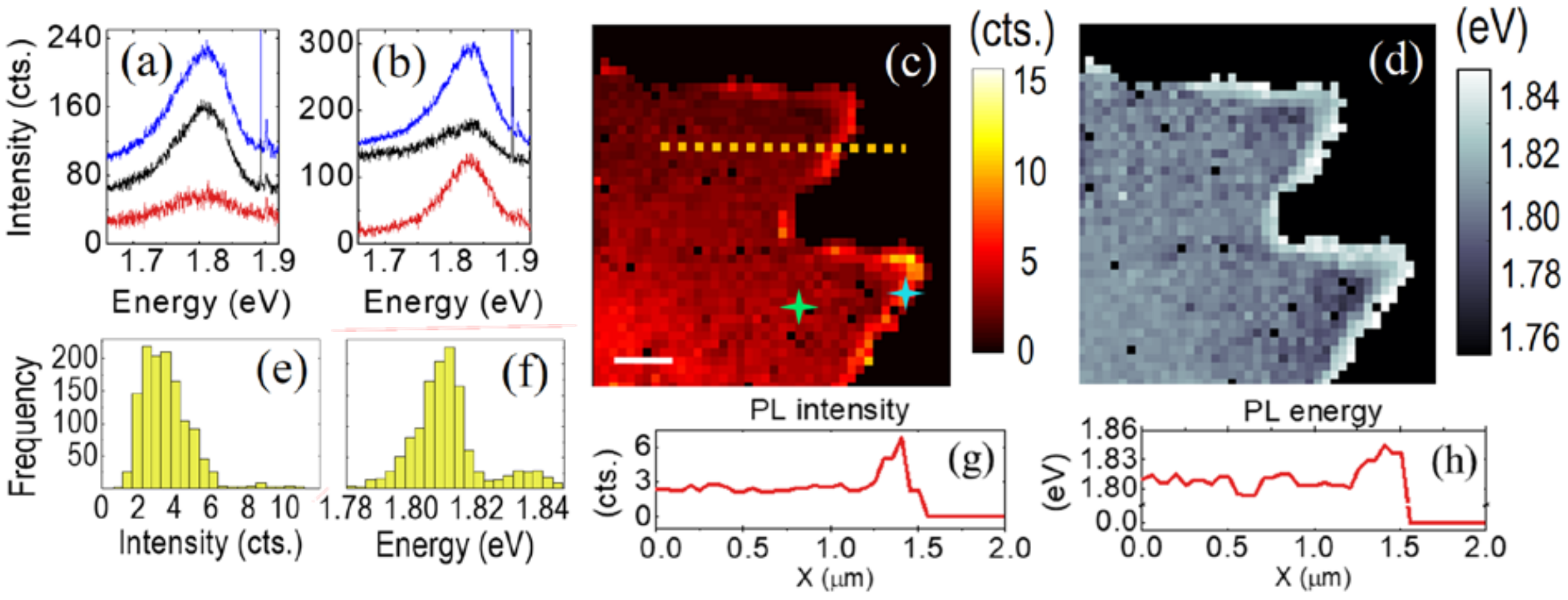

**Figure 2.** TEPL of CVD $MoS_2$ monolayer, the same flake as in Figure 1: PL spectra in the uncorrected near-field (blue line), far-field (black line), and near-field PL (red line) collected (a) in the interior area and (b) at the edge of the flake. The sharp peaks in the spectra are Raman bands of silicon and $MoS_2$ (from left to right). Near-field PL (c) intensity and (d) energy maps; pixel size is 70 nm. Green and blue marks indicate the spots where the spectra in (a) and (b) were collected, respectively. Histograms of near-field PL (e) intensity and (f) energy from the whole map. (g) PL intensity and (h) PL energy profiles across the flake edge in the respective near-field maps. Scale bar in (c) is 500 nm.

Figure 2 reveals two clear conclusions: (i) even in the gap-less mode, TEPL on single-layer $MoS_2$ on a dielectric substrate can provide high enough enhancement to conduct hyper-spectral mapping with lateral resolution down to tens of nanometers, allowing resolution of very detailed features compared to the far-field imaging, and (ii) the resolved PL is clearly more intense at the edges of the flake and more blue-shifted than the flake interior. We extracted both the intensity and energy profiles from the near-field image across the flake edge (Figure 2g and 2h). The profiles show unambiguously the correlated abrupt increase in the PL intensity and energy approaching the edge. The intensity increase at the edge is almost two-fold, and the energy is blue-shifted by approximately 25 meV. This can also be observed in the histograms presented in Figure 2e and 2f. There are two distinct maxima in the PL energy distribution, one centered at 1.809±0.006 eV for the flake interior and the second at 1.835±0.005 eV for the edges (Figure 2e).

TEPLs of two additional, different $MoS_2$ flakes were measured to verify the universality of the edge-related changes in our samples. The obtained AFM images and TEPL data (maps, profiles, and energy histograms) can be found in the Supplementary data (**Figure S2 and S3**). Besides the slight overall shift of the PL distribution, the results are very similar to the previous case. The flake with hexagonal shape (Figure S2) shows the average PL energy values of 1.833±0.005 eV for the interior and 1.845±0.004 eV for the edges, and the average PL energy values for the triangular flake (Figure S3) are 1.819±0.003 eV in the interior and 1.839±0.004 eV at the edges. In addition to the PL results, one can easily detect that the PL distribution correlates very well with the adhesion and surface potential (CPD).

The PL blue-shift and intensity increase are strongly reminiscent of the effect of the decreasing population of charged excitons—the trions—by an electric field,[16] molecular doping,[17] or defects.[7a] As the excessive charges are neutralized, the lower energy trion PL peak becomes weaker, and, at the same time, the higher energy neutral exciton PL peak becomes significantly stronger. The trion–exciton energy difference (i.e. the trion binding energy) varies around 40 meV, depending on the carrier concentration, when manipulated by external bias.[18] As the PL spectral weight gradually moves from the prevailing trion emission at the higher electron concentration to the neutral exciton emission when the specimen gets closer to the charge neutrality point, the maximum PL energy increases from ~1.81 to 1.85 eV.[18] Such a PL energy shift can also be observed in our CVD samples when moving from the flake interior toward its edge. The same effect can be facilitated by defects; for example, Tongay et al.[7a] reported a three-fold increase in the PL integrated intensity as well as a blue-shift of the PL maxima by 20 meV caused by α particle irradiation in the presence of electron-withdrawing $N_2$ molecules, which are bound to the formed defect sites. A higher abundance of sulfur vacancy sites at the edges was reported in several works (e.g. in ref [19]). Oxygen or water molecules bound to the defect sites then induce similar PL enhancement and shift,[17, 20] accompanied by a surface potential shift of up to 150 meV.[19b] Kim et al.[21] ascribed the PL intensity and energy increase at the edges to an enhanced formation of many-body excitons with the aid of charge carriers associated with the defects.

The defect occurrence at the edges connected with electron depletion is universal in our type of growth. The evidenced PL intensity increase at the edges is, actually, the inverse of a previous study that was conducted with TEPL on as-grown $MoS_2$ from an $MoO_3$ precursor.[4i] Decreased PL intensity at the edges was also observed by Lee et al. using optical near-field PL imaging on transferred CVD $MoS_2$, and the low PL yield was ascribed to structural disorder.[22] The diversity of defect-induced PL modification in CVD TMDCs is clearly very large, ranging from simple non-radiative recombination of excitons at the defect sites to enhanced biexciton formation or to suppressed trion formation, depending on the carrier density associated with the defects. It can be assumed that there should be additional factor(s) controlling the excitonic behavior of the defects.

Ly et al. suggested that PL enhancement at the edges could be a consequence of the mechanical buckling initiated by delamination at the edges producing sites with different chemical activity.[23] We measured, using a sharper, non-silver-coated AFM tip, the nanomechanical properties of our flakes focusing on the edge to determine the sources of topography increase shown in Figure 1, S2, and S3, and to establish whether there is any evidence of delamination at the edges. **Figure 3a, 3b, and 3c** show the topography, deformation, and adhesion images, respectively, of an edge of the CVD monolayer $MoS_2$ flake. At first glance, it is clear that some of the nanoparticles exhibit different adhesion and deformation contrasts. Moreover, the heterogeneous region close to the edge and running roughly parallel to it also shows different deformation and adhesion properties. More detailed, zoomed-in images are shown in Figure 3d, 3e, and 3f, in which the two types of nanoparticles can be clearly distinguished. It appears that the particles that present a diffuse contrast in the topography image are below the $MoS_2$ layer (red arrows in Figure 3d), which can be verified upon closer inspection of the adhesion and deformation images. Additional images of different flakes and their edges can be found in the Supplementary Information (**Figure S4**).

This observation is an indication of partial delamination taking place at the edges of the flakes, possibly due to strain induced during the cooling stage after the growth,[23] as has been well described in graphene.[24] The delamination is not full and is aided by the particles that promote and accompany the $MoS_2$ layer growth.[12] When the particles are trapped under the $MoS_2$ edge at the end of the growth process, they induce buckling and delamination of the layer from the substrate. It has been well documented that the Si/$SiO_2$ substrate induces n-doping of 1L $MoS_2$, which in turn causes PL intensity to decrease and a red shift to ~1.82 eV (i.e. the trion prevalence), while free-standing $MoS_2$ exhibits blue-shifted and more intense PL due to neutral exciton formation.[25] The estimated minimum Fermi level energy shift induced by the substrate doping was at least 40 meV.[25b] The evolution of the PL behavior from the interior to the edge of our CVD flakes very closely resembled the reported supported–free-standing $MoS_2$ PL difference. Moreover, the measured surface potentials, whose values were in the order of substrate > flake interior > flake edge, further corroborate the important role of the substrate beyond its role in the defects that originate in the imperfect growth at the edges. In capacitance scanning, the contrast in semiconductors is provided by the dopant (charge) nature and concentration, while thick dielectric films show very small capacitance. Hence the $\partial^2C/\partial z^2$ profile from the interior of the flake to the $SiO_2$ substrate (Figure 1i) shows decreasing $\partial^2C/\partial z^2$ values. The observed carrier depletion at the edges is in accordance with the other obtained results (CPD, TEPL). We also note the possibility of a potential contribution of additional effects arising from the corrugated topography that has been observed in GaSe layers deposited on graphite, in which the increase in sample volume and the changed crystal orientation at the deformed edges were conjectured to be responsible for the observed PL intensity increase.[26]

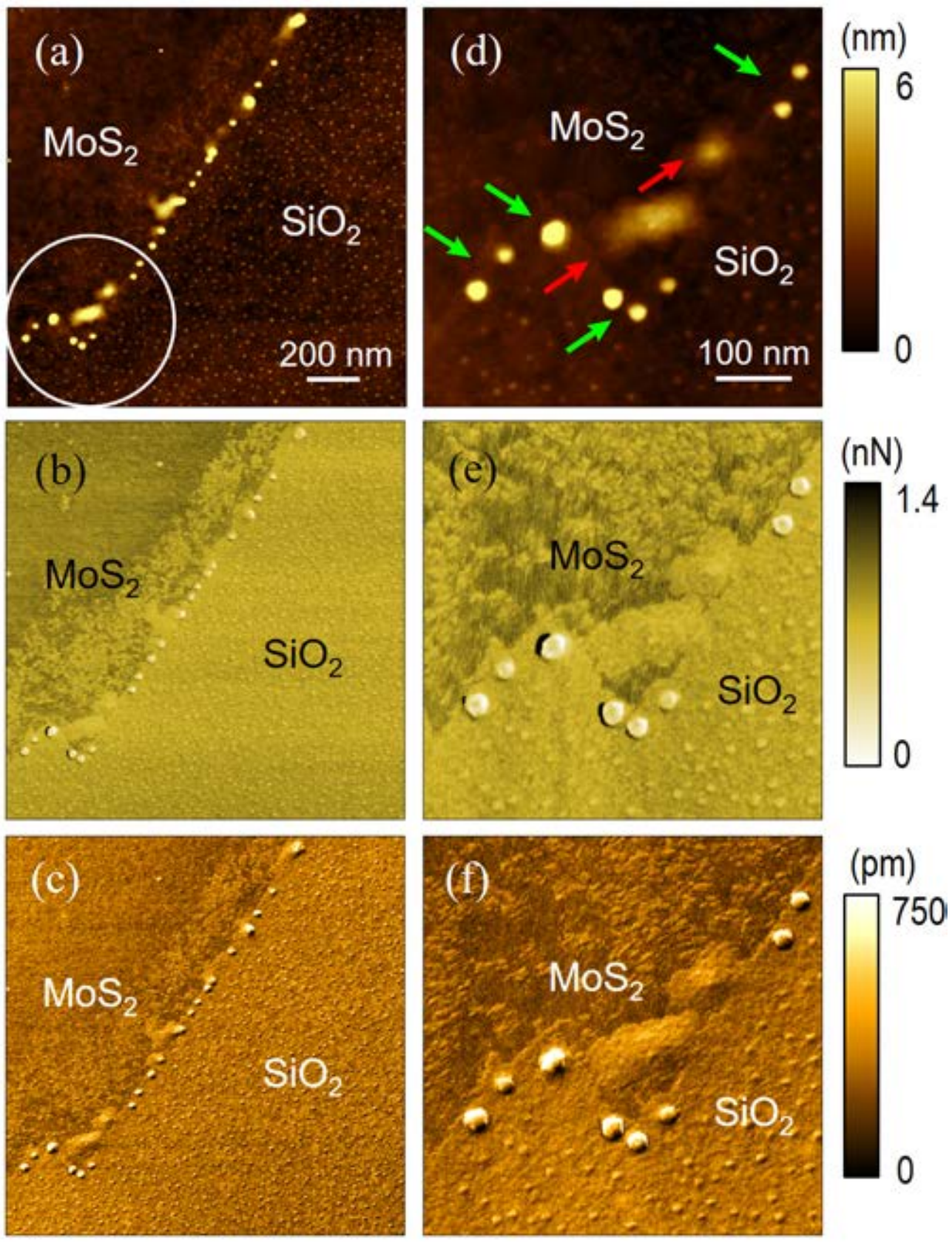

**Figure 3.** Non-quantitative nanomechanical properties of CVD-grown $MoS_2$ monolayer; (a) topography, (b) adhesion, and (c) deformation maps. An enlarged area indicated with a circle in (a) is shown in images (d)–(f). Red and green arrows in (d) show covered and free segregation nanoparticles, respectively. Adhesion and deformation range minima were set to zero.

We successfully applied gap-less TEPL to characterize as-grown CVD $MoS_2$ flakes. To contrast with our observations, we next focused on a mechanically exfoliated flake. The topography image in **Figure 4a** shows a flake formed by different numbers of layers (increasing from the edge on the right). Additional AFM images are shown in **Figure S5**. A large wrinkle (100 nm wide, 15 nm high) through the whole length in the middle of the image and two smaller ones approximately perpendicular to the edge can be observed in the monolayer region. These features and the separation between the layers are better observed in both the CPD and $\partial^2C/\partial z^2$ magnitude images (**Figure S5**), which provide, qualitatively, the same information.

In Figure 4b, 4c, 4d, and 4e, the far- and near-field PL intensity and energy maps of the exfoliated flake are shown for comparison. Both the large and small wrinkles are distinguished in the near-field maps. Selected spectra are shown in Figure 4f and 4g. The spectra obtained from the large wrinkle (Figure 4g, marked blue in Figure 4a) show a slightly higher enhancement than the flat monolayer region (Figure 4f, marked green in Figure 4a). From the maps, we can conclude that the three wrinkles exhibit higher PL intensity than the rest of the flake as well as shifts in the energy. Curiously, the large wrinkle red-shifts the PL energy, which is counterintuitive, since small compressive strain should cause band-gap opening, i.e. PL energy increase.[27] In a similar case, albeit with few $MoS_2$ layers and much wider wrinkles, the unexpected red shift was explained by the effects of energy funneling.[28] In contrast, the PL energy at the small wrinkles is blue-shifted, and so the compressive effects probably dominate. The increased PL energy can be again, at least in part, understood through the combination of the increased sample volume and delamination from the substrate, which causes de-doping and thereby favors the neutral exciton transition, similar to the edges in CVD graphene. However, this seems to be fully valid only for the smaller wrinkles, which show the expected energy blue-shift.

The most important observation with respect to the CVD samples is the rather uniform PL in terms of both intensity and energy going from the interior of the flake toward the edge. No tonality at the edge can be observed in the CPD and $\partial^2C/\partial z^2$ magnitude images either (Figure S5), confirming that the defect/delamination front at the edges in the CVD samples is inherent to the growth process and not due to, for example, aging of the sample in laboratory conditions.

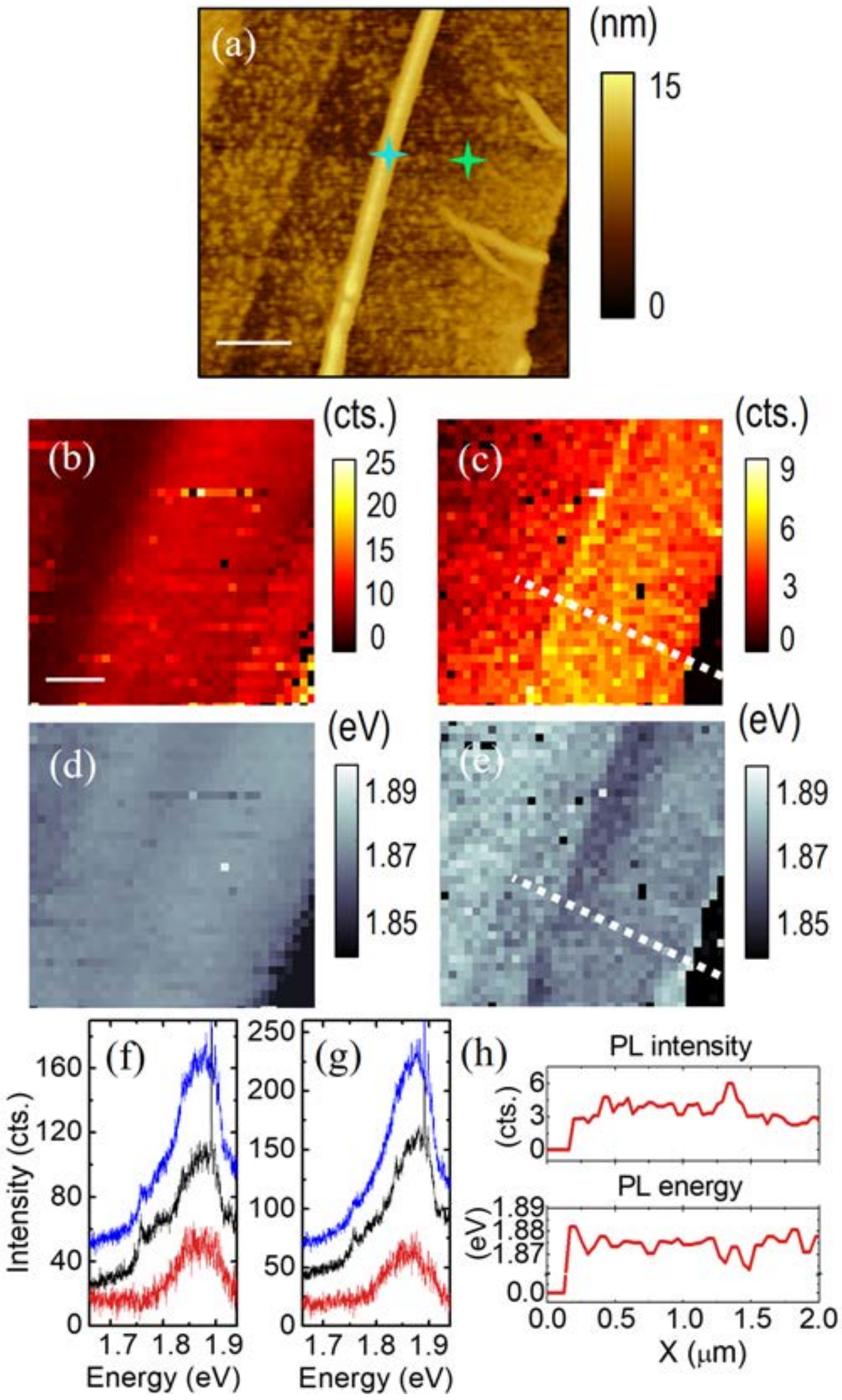

**Figure 4.** TEPL of exfoliated $MoS_2$ flake. (a) AFM topography image. Green and blue marks indicate the areas where the spectra in (f) and (g) were collected, respectively. (b) Far- and (c) near-field PL intensity maps and (d) far- and (e) near-field PL energy maps. Pixel size is 70 nm. PL spectra in the uncorrected near-field (blue line), far-field (black line), and near-field (red line) collected at (f) the “flat” area of the flake and (g) the large wrinkle. (h) PL intensity and energy profiles along near-field maps. Scale bar is 500 nm.

In conclusion, we have shown that gap-less TEPL on monolayer $MoS_2$ is capable of providing signals intense enough to thoroughly investigate as-grown samples on a dielectric substrate with a lateral resolution down to tens of nanometers. Consequently, there is no need to transfer the samples to a metallic (gold or silver) substrate before such an analysis takes place, thereby avoiding modification (or even damage) and contamination of the sample by the transfer process. Fitting the spectra in every map point reveals that monolayer $MoS_2$ flakes exhibit heterogeneities in samples grown by CVD and prepared by mechanical exfoliation. The edges of the CVD flakes feature, in all the samples, an increased intensity and a blue-shift of the PL peak originating from the A exciton. The PL results correlate well with surface potential and differential tip-sample capacitance measurements, which manifest values clearly distinguished between the flake edges and interiors. With the help of additional nanoscale-resolution data obtained by nanomechanical AFM mapping, the changes in the PL signatures and in the surface potential can be explained by the combination of growth-induced defects and the absence of substrate doping effects due to partial delamination of the $MoS_2$ edges. In contrast, the interior of the CVD flakes is rather homogeneous in the single-layer limit. Both these observations are in stark contrast with measurements of exfoliated layers. In this case, the edges have similar PL intensity and energy to the flake interior. The variations in PL signature are clearly visible in the whole flake, caused by topographical corrugations, like wrinkles or bubbles, and contamination.

**Experimental Section**

*Sample preparation*

The $MoS_2$ monolayers were grown using $MoO_2$ (Sigma Aldrich #234761) and S (Sigma Aldrich #344621) as sources by atmospheric pressure CVD.[4g, 29] A 3 × 1 cm piece of $Si/SiO_2$ (300 nm) substrate was cleaned with acetone and isopropanol and subsequently placed face-down on a quartz crucible containing 30 mg $MoO_2$ powder. The crucible was installed in the middle of the oven, and 90 mg S was placed just outside the oven. The tube was flushed with Ar gas at room temperature, and the $MoS_2$ growth was initiated with an Ar flow of 120 sccm. The temperature was gradually increased to the growth temperature of 1123 K at 40 K/min and held there for 10 minutes. After the growth, the oven was opened, and the sample was quickly cooled to room temperature in the Ar flow of 120 sccm.

The $MoS_2$ exfoliated layer was prepared from $MoS_2$ bulk (HQ graphene) by mechanical exfoliation onto polydimethylsiloxane (PDMS) and transferred with a customized transfer station on top of an $SiO_2$/Si substrate previously cleaned with acetone and isopropanol.

*AFM and Raman details*

Topography in tapping mode and CPD, together with $\partial^2C/\partial z^2$ magnitude images obtained by KPFM-FM, were obtained with an AIST-NT SPM, using Pt-Ir–covered tips (ACCESS-EFM probes, AppNano, k = 2.7 $N.m^{-1}$, f = 60 kHz). Topography and mechanical property images were measured with an Icon Dimension AFM (Bruker) using PeakForce tapping mode. Scanasyst-Air Bruker probes were used in this case. Micro Raman and PL spectra were collected with a LabRAM Evolution spectrometer using 532 nm laser excitation with 1800 and 300 l/mm gratings, respectively. TEPL measurements were performed with a LabRAM HR Evolution spectrometer coupled to an AIST-NT SPM with 633 nm laser excitation, p-polarization, and 300 l/mm grating, using side optical access. The integration time for each pixel was 0.25–0.5 sec, depending on the number of pixels, with a laser power below 300 μW. Silicon ACCESS-FM probes (App Nano, tip/view, k = 2.7 $N.m^{-1}$, f = 60 kHz) were covered by approximately 70 nm of silver by magnetron sputtering.

**Supporting Information**
Supporting Information is available from the Wiley Online Library or from the author.

**Acknowledgements**
This work was funded by Czech Science Foundation (GACR 17-18702S). A.R gratefully acknowledges the financial support from the Czech Republic Ministry of Education, Youth and Sport through Project Nr. CZ.02.2.69/0.0/16_027/0008355. T.V. and J.V. acknowledge the European Research Council (ERC-Stg-2016 TSuNAMI, project no. 716265). The work was further supported by the project Pro-NanoEnviCz (Reg. No. CZ.02.1.01/0.0/0.0/16_013/0001821) supported by the Ministry of Education, Youth and Sports of the Czech Republic and the European Union - European Structural and Investments Funds in the frame of Operational Programme Research Development and Education.

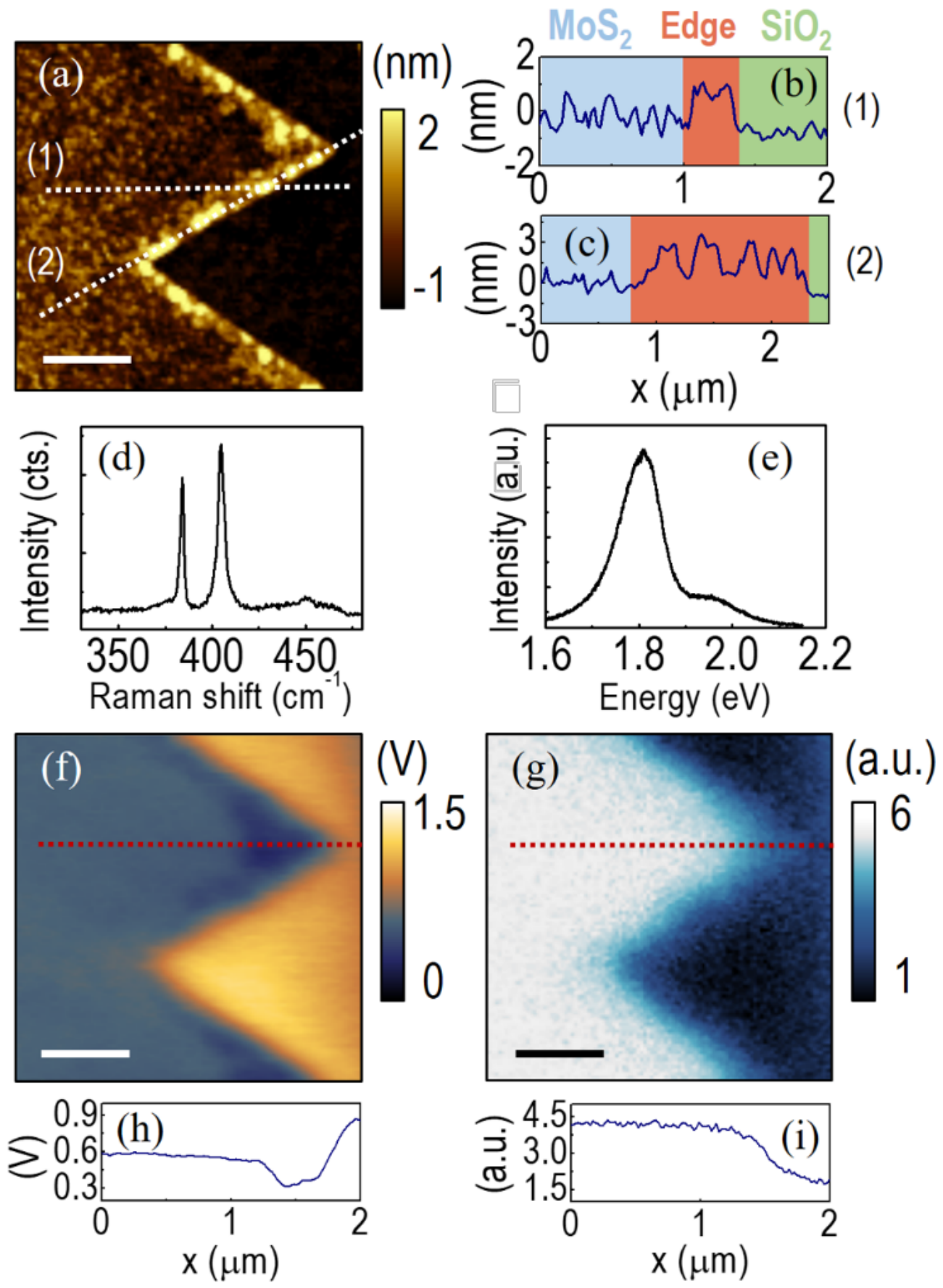

**Figure 1.** CVD-grown $MoS_2$ monolayer: (a) topography image and height profiles (b) across and (c) along the edge. (d) Raman spectrum and (e) photoluminescence spectra acquired in the flake interior. (f) CPD and (g) $\partial^2C/\partial z^2$ maps of the same area with (h) CPD and (i) $\partial^2C/\partial z^2$ profiles across the edge. Scale bar in each map is 500 nm.

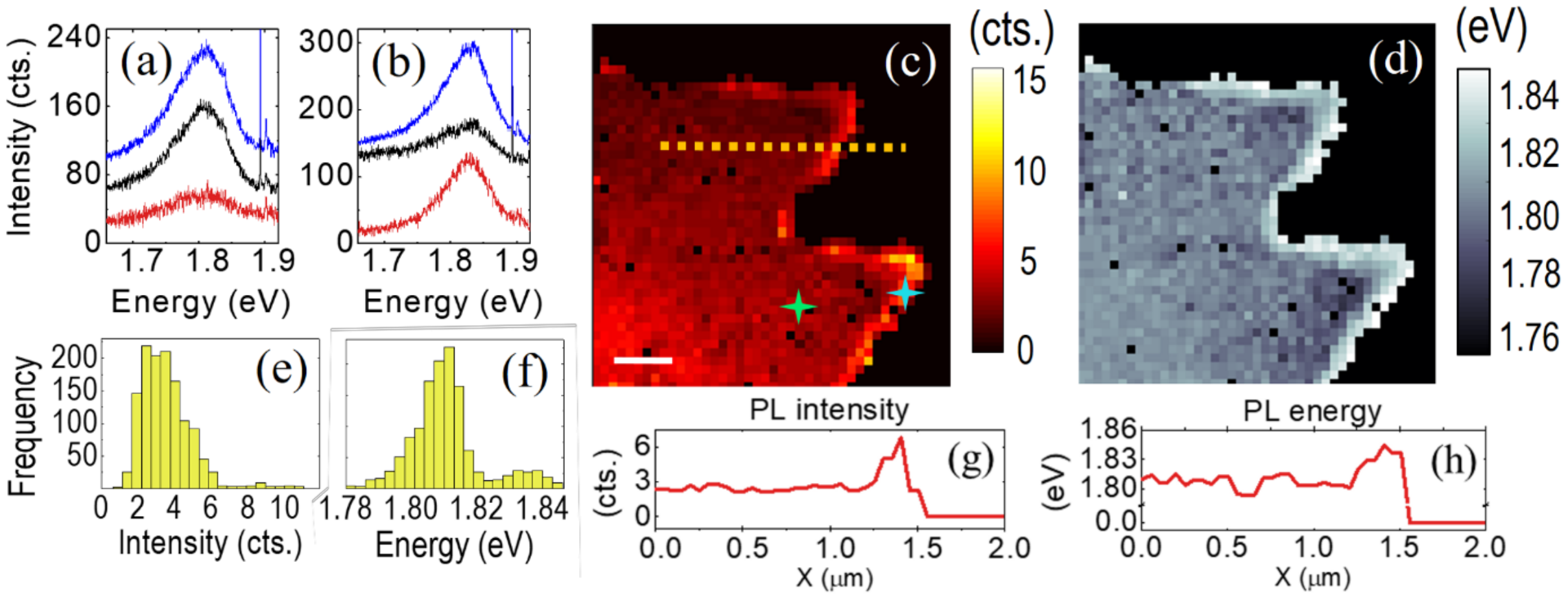

**Figure 2.** TEPL of CVD $MoS_2$ monolayer, the same flake as in Figure 1: PL spectra in the uncorrected near-field (blue line), far-field (black line), and near-field PL (red line) collected (a) in the interior area and (b) at the edge of the flake. The sharp peaks in the spectra are Raman bands of silicon and $MoS_2$ (from left to right). Near-field PL (c) intensity and (d) energy maps; pixel size is 70 nm. Green and blue marks indicate the spots where the spectra in (a) and (b) were collected, respectively. Histograms of near-field PL (e) intensity and (f) energy from the whole map. (g) PL intensity and (h) PL energy profiles across the flake edge in the respective near-field maps. Scale bar in (c) is 500 nm.

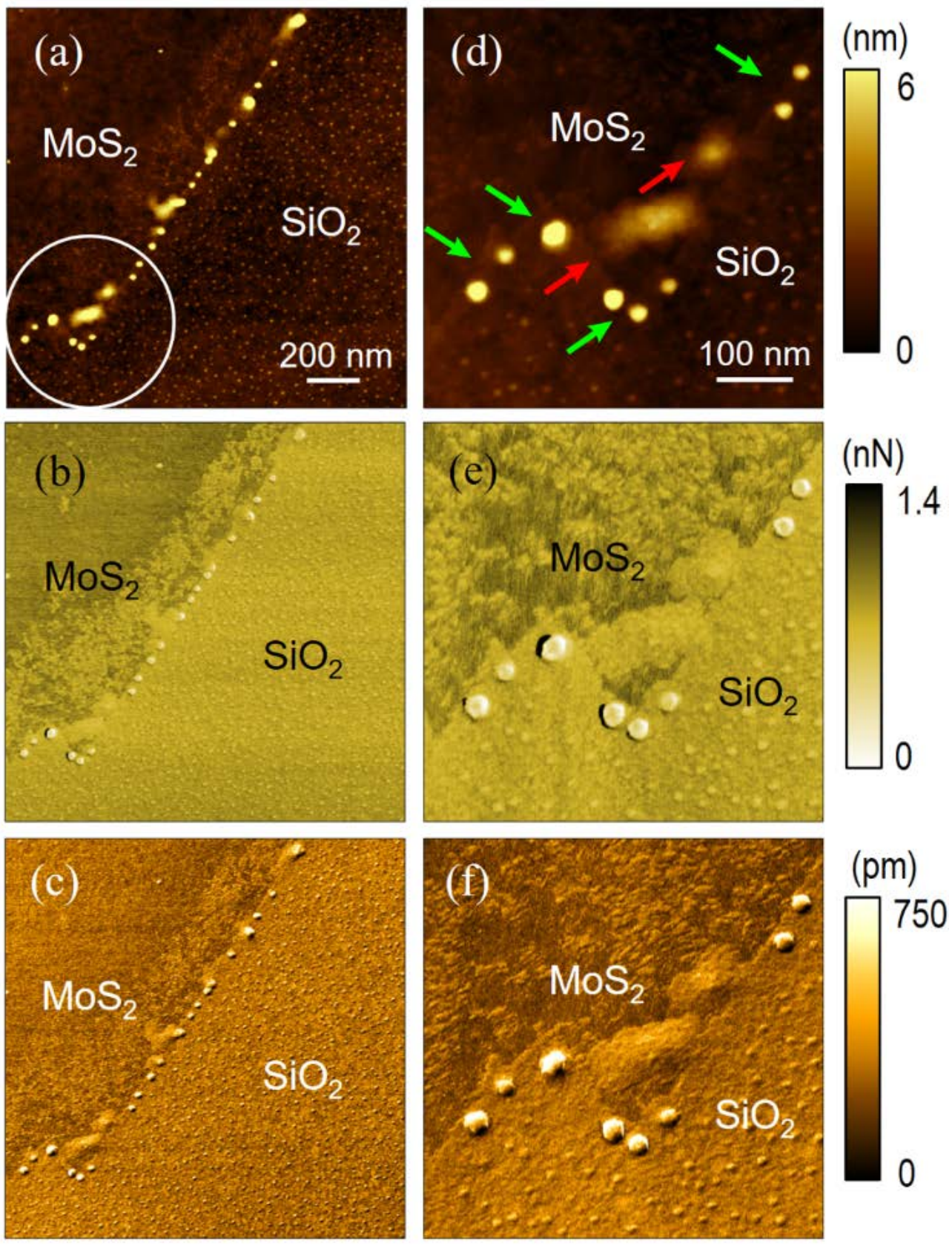

**Figure 3.** Non-quantitative nanomechanical properties of CVD-grown $MoS_2$ monolayer; (a) topography, (b) adhesion, and (c) deformation maps. An enlarged area indicated with a circle in (a) is shown in images (d)–(f). Red and green arrows in (d) show covered and free segregation nanoparticles, respectively. Adhesion and deformation range minima were set to zero.

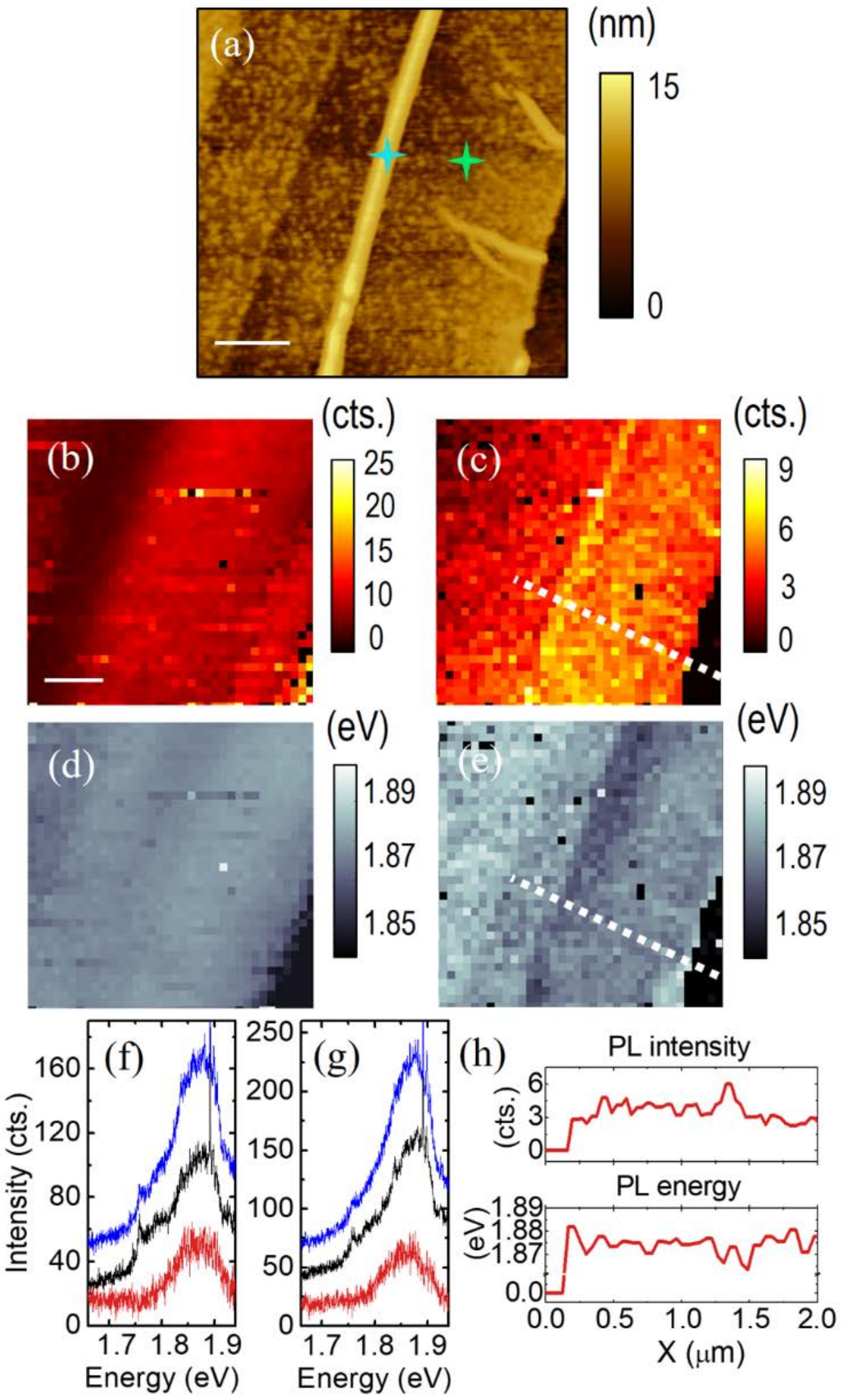

**Figure 4.** TEPL of exfoliated $MoS_2$ flake. (a) AFM topography image. Green and blue marks indicate the areas where the spectra in (f) and (g) were collected, respectively. (b) Far- and (c) near-field PL intensity maps and (d) far- and (e) near-field PL energy maps. Pixel size is 70 nm. PL spectra in the uncorrected near-field (blue line), far-field (black line), and near-field (red line) collected at (f) the "flat" area of the flake and (g) the large wrinkle. (h) PL intensity and energy profiles along near-field maps. Scale bar is 500 nm.

Tip-Enhanced Photoluminescence in the gap-less mode is utilized to discern local (tens of nm) inhomogeneities in the optoelectronic properties of as-grown $MoS_2$ monolayers. The results are put in contrast to measurements on exfoliated $MoS_2$. Special consideration is given to the edges, where photoluminescence enhancement and shift are observed in the grown $MoS_2$ and ascribed to the edge delamination.

**Tip-Enhanced Spectroscopy**

A. Rodriguez, T. Verhagen, M. Kalbac, J. Vejpravova, O. Frank*

**Imaging Nanoscale Inhomogeneities in As-Grown $MoS_2$ by Tip-Enhanced Photoluminescence in Gap-Less Mode**

ToC figure

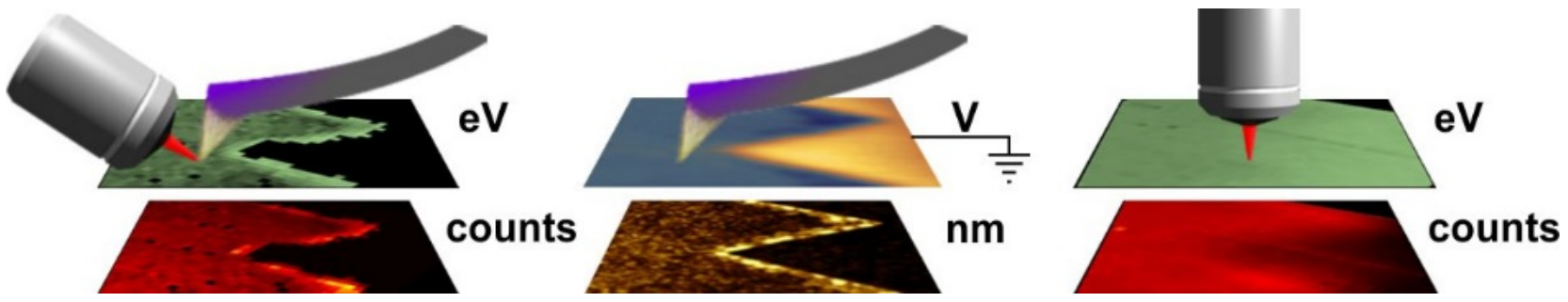

# Supporting Information

**Imaging Nanoscale Inhomogeneities and Edge Delamination in As-Grown $MoS_2$ Using Tip-Enhanced Photoluminescence**

*Alvaro Rodriguez, Tim Verhagen, Martin Kalbac, Jana Vejpravova, Otakar Frank**

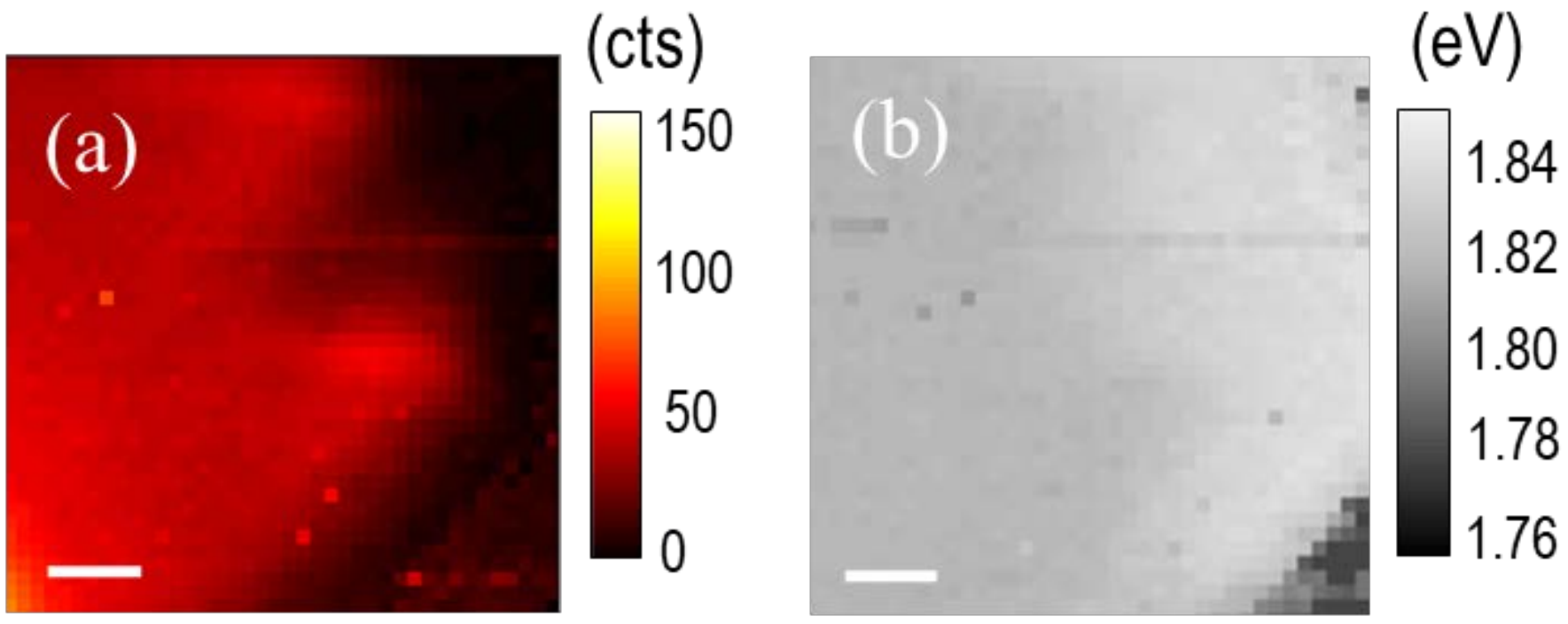

Figure S1. Far-field PL intensity and energy maps of CVD grown $MoS_2$ monolayer. Pixel size is 70 nm. Scale bar is 500 nm.

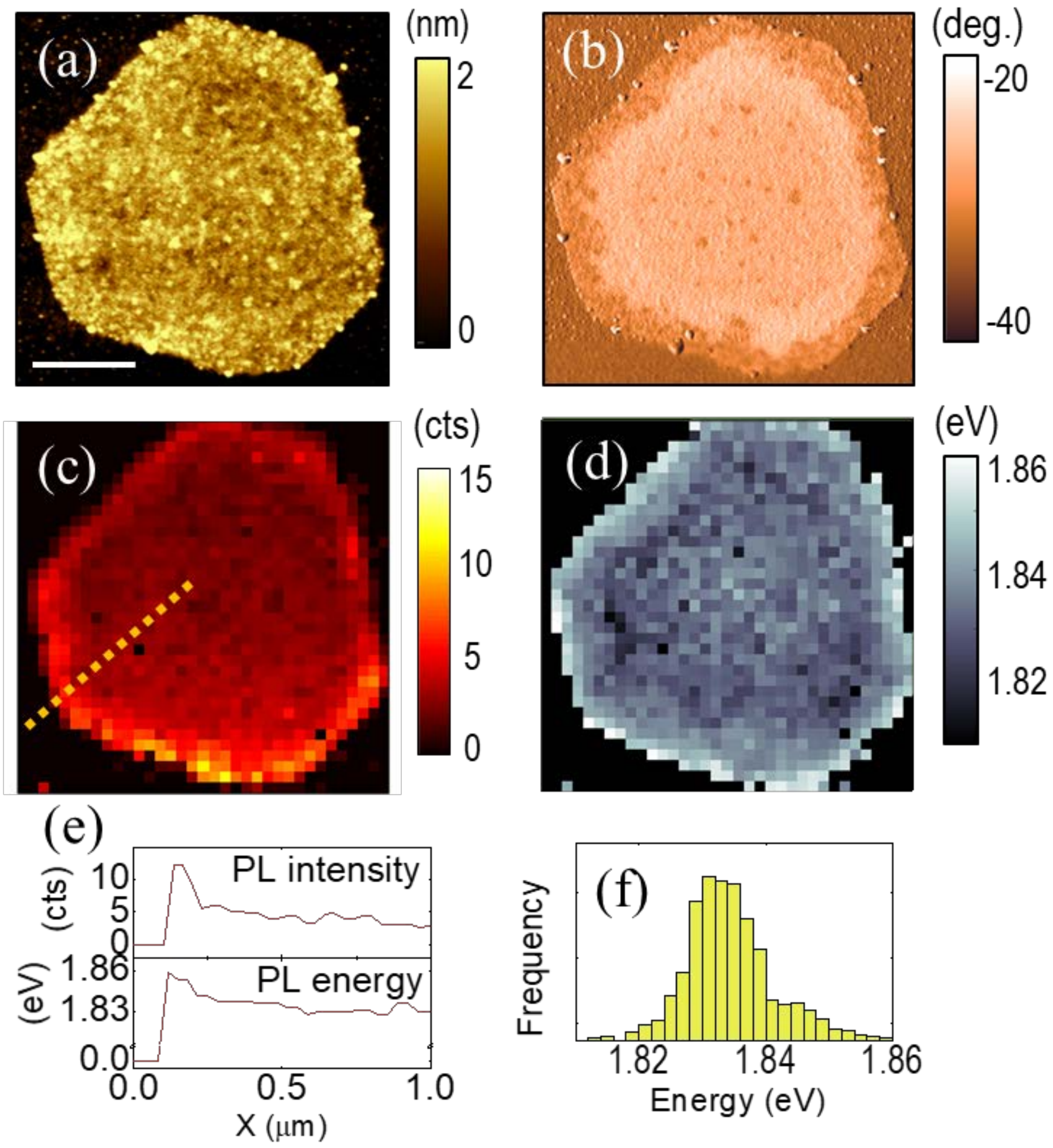

Figure S2. CVD grown $MoS_2$ monolayer with hexagonal shape. (a) Topography AFM image measured in tapping mode and (b) phase image. (c) Near-field TEPL intensity and (d) energy maps. Pixel sized is 60 nm. (e) PL intensity and energy profiles across the flake edge for the near-field map, (f) histogram of near-field PL energy obtained from the map. Scale bar is 500 nm.

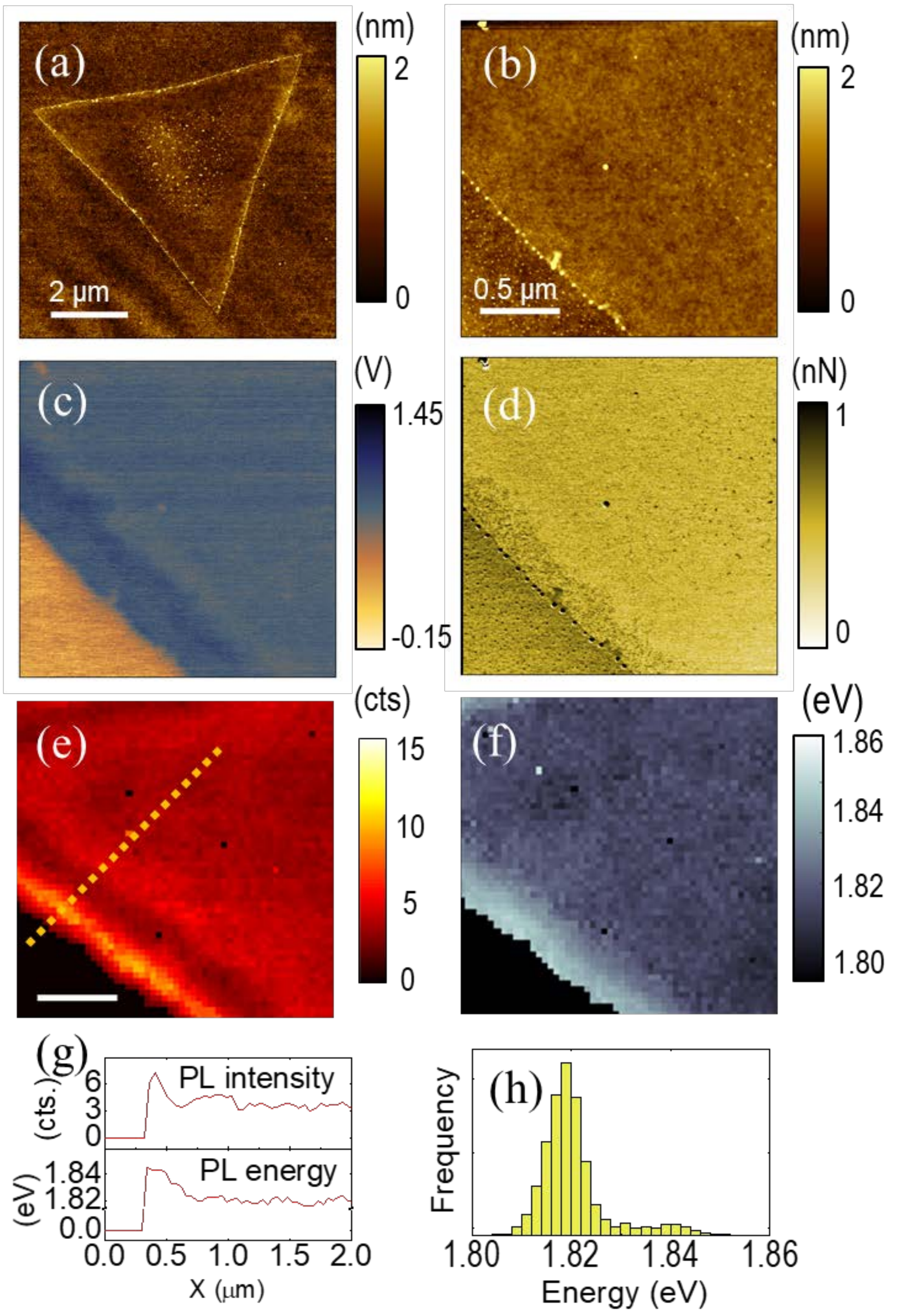

Figure S3. CVD grown $MoS_2$ monolayer with triangular shape. (a) Topography AFM image measured in peak force tapping mode and (b) enlarged area focusing on the edge, (c) surface potential image measured in Peak Force KPFM and (d) adhesion image. (e) Near-field TEPL intensity and (f) energy maps. Pixel size is 30 nm. (g) PL intensity and energy profiles across the flake edge for the near-field map, (h) histogram of near-field PL energy obtained from the map. Scale bar is 500 nm.

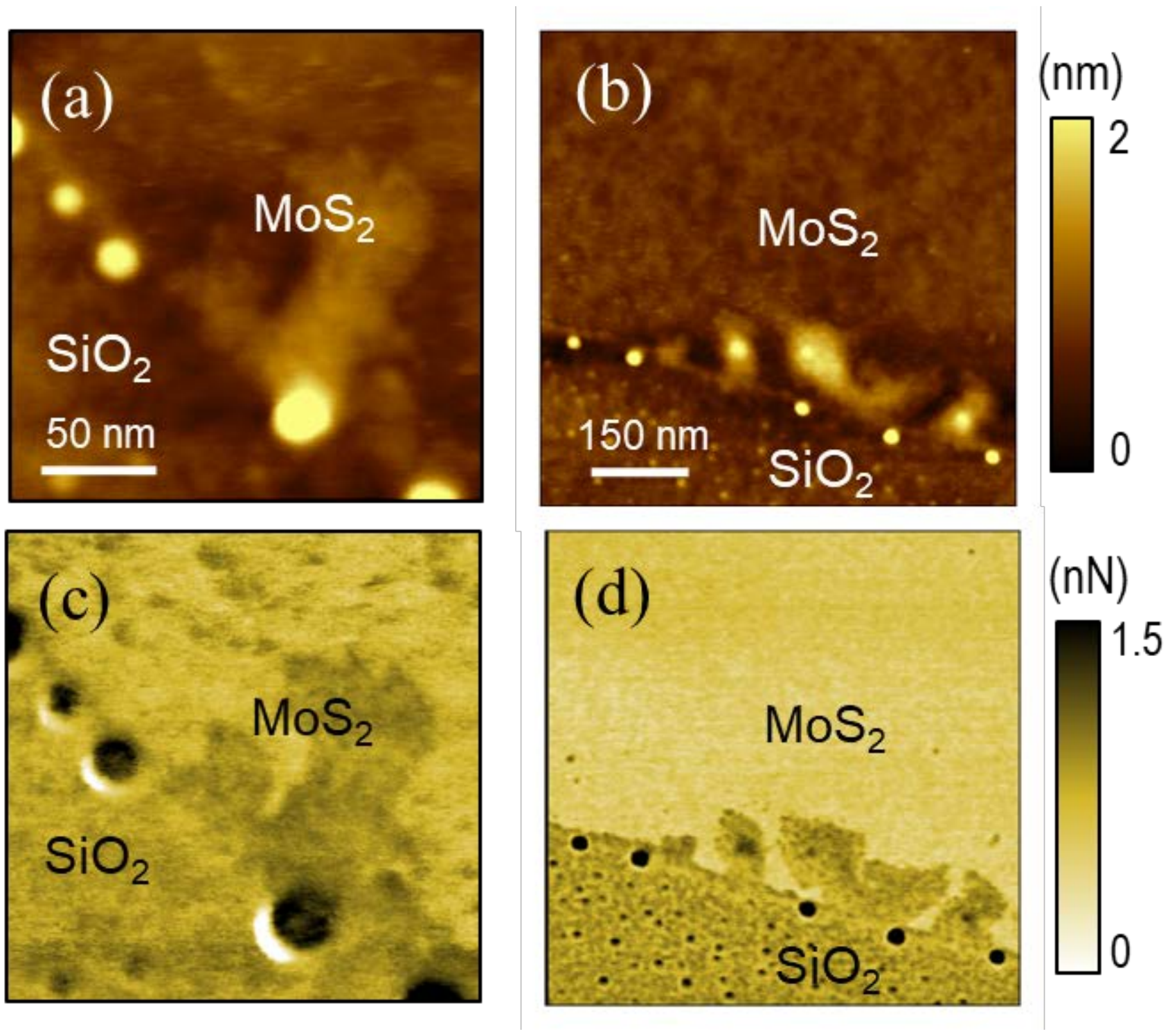

Figure S4. AFM images showing particles below the CVD grown $MoS_2$ monolayer. (a-b) Topography AFM images and (c-d) adhesion images.

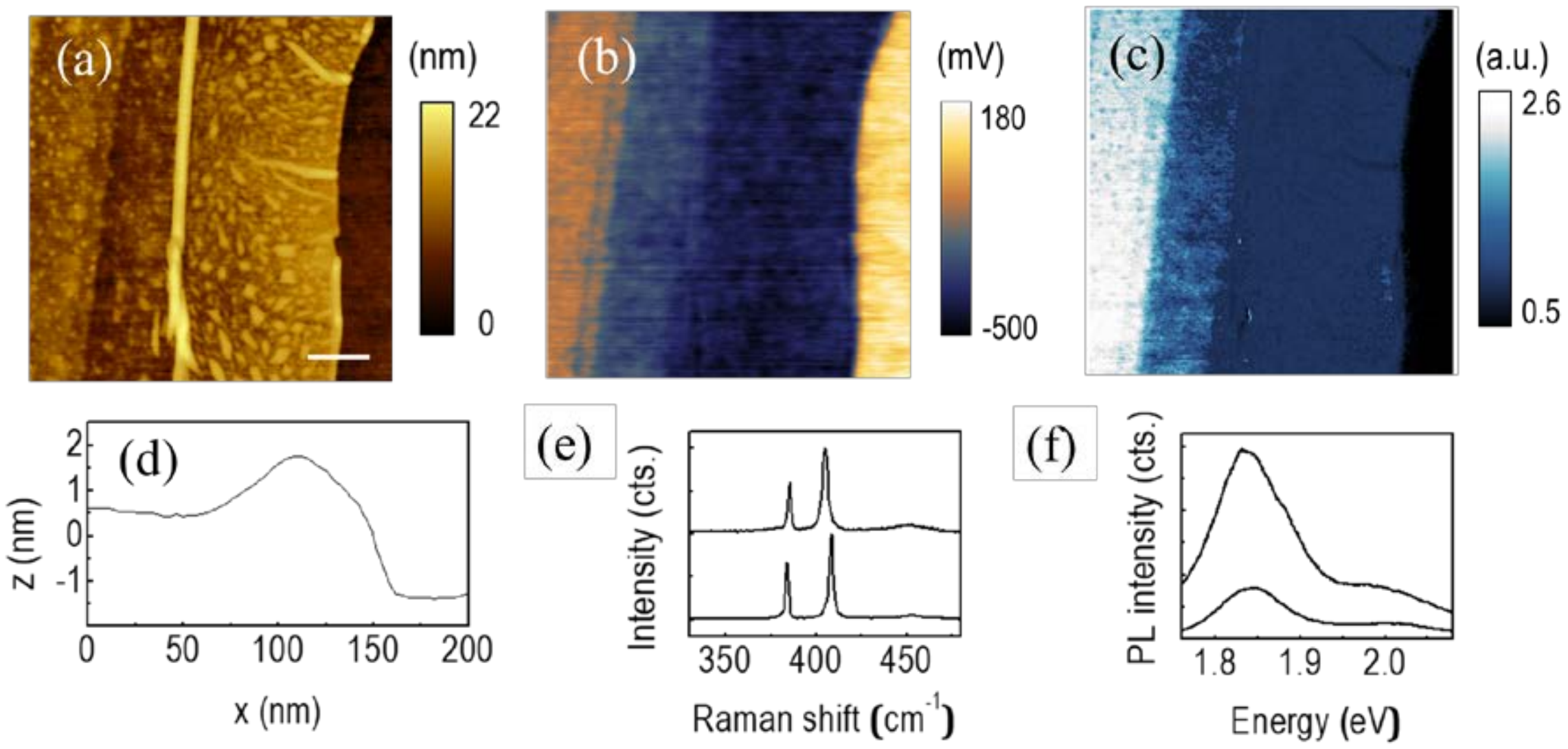

Figure S5. Exfoliated $MoS_2$ flake: (a) AFM topography, (b) CPD image and (c) $\partial^2C/\partial z^2$ images measured with KPFM-FM. (d) Height profile, (e) Raman spectra of the monolayer (top) and few layers (bottom), (f) PL spectra of the monolayer (top) and few layers (bottom). Scale bar is 500 nm.

The reason of the large thickness observed by AFM in the exfoliated monolayer has been previously reported [1], it was observed that 1 nm of PDMS remains after the transfer. Raman spectra confirm the monolayer thickness. [2,3].

References (Supporting Information).